\documentclass[a4paper,11pt]{article}
\usepackage{pos}
\usepackage{comment}

\title{Control variates with neural networks}

\author*{Hyunwoo Oh}

\affiliation{Department of Physics, University of Maryland,\\
College Park, MD 20742, USA}

\emailAdd{hyunwooh@umd.edu}

\abstract{ 

The precision of lattice QCD calculations is often hindered by the stochastic noise inherent in these methods. The control variates method can provide an effective noise reduction but are typically constructed using heuristic approaches, which may be inadequate for complex theories. In this work, we introduce a neural network-based framework for parametrizing control variates, eliminating the reliance on manual guesswork. Using $1+1$ dimensional scalar field theory as a test case, we demonstrate significant variance reduction, particularly in the strong coupling regime. Furthermore, we extend this approach to gauge theories, showcasing its potential to tackle signal-to-noise problems in diverse lattice QCD applications.
}

\FullConference{The 41st International Symposium on Lattice Field Theory (LATTICE2024)\\
28 July - 3 August 2024\\
Liverpool, UK\\}

\begin{document}
\maketitle

\section{Introduction}

Lattice Quantum Chromodynamics (QCD) using Monte Carlo methods has achieved remarkable success across a wide range of problems. However, these methods face significant limitations, particularly when dealing with large baryon chemical potentials or real-time dynamics, due to the well-known sign problem. Even in the absence of a baryon chemical potential and under thermodynamic equilibrium, certain observables are plagued by the signal-to-noise problem, which can render Monte Carlo calculations exponentially challenging.

A novel approach to reducing variance is by the construction of a control variate~\cite{CV}. The idea of control variates is simple. If an observable $O$ is given, an unbiased estimator $O-f$ can be constructed, provided that $f$ satisfies $\langle f \rangle = 0$. However, their variances can be different:
\begin{equation}
    \langle (O -f)^2 \rangle = \langle  O^2 \rangle + \langle f^2 \rangle - 2 \langle  O f \rangle. \label{Eq:CV}
\end{equation}
which implies ${\rm Var}(O-f) = {\rm Var}(O) - 2 \langle  O f \rangle +  \langle f^2 \rangle$.
Therefore, finding $f$, referred to as a control variate, that is highly correlated with $O$ and has a small expectation value of its square, can potentially reduce the variance of $O-f$.

In general, identifying an operator with a zero expectation value, ensuring that the new observable $\tilde{O} \equiv O - f$ remains unbiased, is a challenging task. One way to find such operators are through the Stein's method~\cite{Stein} (or so-called the lattice version of the Schwinger-Dyson equations~\cite{CV, FCV}):
\begin{equation}
    \int D\phi \frac{\delta}{\delta \phi} (g {\rm e}^{-S}) =0 \label{Eq:Stein} 
\end{equation}
for a well-defined $g$ with a proper boundary condition.
Therefore, if a control variate $f$ is defined as $ f \equiv \partial g - g \partial S$, it directly follows that $\langle f \rangle = 0$. The objective then becomes finding a function $g$ which minimizes Eq.~(\ref{Eq:CV}). In this work, we study the control variates method with neural network parametrizations, by summarizing and extending the work in~\cite{Oh}.


\section{Neural network}

While there are attempts to find an optimal $g$ from perturbative ways~\cite{CV,FCV}, one can directly parametrize g with a neural network. Neural networks provide a framework for parametrizing functions through a sequence of affine transformations followed by activation functions. Since neural networks are known to be universal approximators~\cite{MLP, MLP2}, they offer a promising approach for identifying more effective choices for $g$.

The general structure of a neural network is as follows:
\begin{equation}
    g(\phi) = W_n \sigma_n( \cdots \sigma_1(W_0 \phi + b_0)) + b_n \label{Eq:NN}
\end{equation}
where $W_i$ and $b_i$ represent affine transformations, and $\sigma_i$ are activation functions which is non-linear in general. In our case, the input $\phi$ corresponds to a single configuration on the lattice. When there is no hidden layer, Eq.~(\ref{Eq:NN}) corresponds to an affine transformation.

\subsection{Symmetry} \label{Sec:Sym}

In principle, neural networks can learn the symmetries of the control variates, which are essential for achieving optimal results. However, because the model is trained only a finite number of times using a limited set of training samples, it is advisable to impose constraints on the neural network or on the construction of the control variates. This ensures that the symmetry is built into the framework, reducing the burden on the neural network to learn it during training.

Among many symmetries, the first example is translational symmetry. If both the action and the observable respect the translational symmetry, the control variate $f$ also must preserve it: 
\begin{equation}
    f(T_y[\phi]) = f(\phi),
\end{equation}
where $T_y$ represents a translation with an amount $y$: $T_y[\phi_x] = \phi_{x+y}$.

In this section, we will use the Schwinger-Dyson control variates from a vector-valued function $g$ with the following form:
\begin{equation}
    f(\phi) = \sum_i \left( \frac{\partial g_i}{\partial \phi_i} - g_i \frac{\partial S}{\partial \phi_i} \right)  \; {\rm where} \; g:\mathbb R^V \rightarrow \mathbb R^V, \label{Eq:CV_Divergence}
\end{equation}
where $V$ denotes the degrees of freedom. Other constructions of the Schwinger-Dyson control variates and their universality will be discussed in Section~\ref{Sec:PCV}.

In order for Eq.~(\ref{Eq:CV_Divergence}) to respect the translational symmetry, $g$ should be translationally covariant (or equivariant):
\begin{equation}
    g(T_y[\phi])_x =  g(\phi)_{x+y}.
\end{equation}
Note that $g(\phi)_x$ denotes the $x$th component of the vector-valued function $g$ at $\phi$.

The translational covariance of $g$ can be enforced by first defining a real-valued function $g_0:\mathbb R^n \rightarrow \mathbb R$, which can be parametrized b a neural network, and then constructing a vector-valued function $g$ by 
\begin{equation}
    g(\phi)_x =  g_0(T_{x} [\phi]). \label{Eq:cov}
\end{equation}
With this construction, one can easily check that the control variate $f$ is translationally invariant:
\begin{equation}
    \begin{aligned}
        f(T_y[\phi]) & = \sum_{x} \left(\frac{\partial g(T_y[\phi])_x}{\partial \phi_{x+y}} -  g(T_y[\phi])_x \frac{\partial S(T_y[\phi])}{\partial \phi_{x+y}} \right) \\
        & = \sum_{x} \left( \frac{\partial g_0(T_{x+y}[\phi])}{\partial \phi_{x+y}} - g_0(T_{x+y}[\phi]) \frac{\partial S(T_{x+y}[\phi])}{\partial \phi_{x+y}} \right) 
        = f(\phi) .
    \end{aligned} 
\end{equation}
Moreover, this construction does not depend on the structure of neural networks.

Another example is the $Z_2$ symmetry. It is satisfied if $g$ is an odd function: $g(-\phi) = -g(\phi)$. To ensure that the neural network in Eq.~(\ref{Eq:NN}) is odd, the affine layers $\phi\rightarrow W_i\phi+b_i$
must have zero biases, i.e., $b_i=0$. Additionally, the activation functions must be odd: $\sigma_i(-x) = - \sigma_i(x)$ for each $i$. Suitable activation functions include ${\rm arcsinh}$ or $\tanh$. In this work, all activation functions are chosen to be ${\rm arcsinh}$, as its range is $(-\infty, \infty)$.

\subsection{Training and overfitting}

The ideal loss function, $L$, for optimizing the performance of control variates is the variance of $\tilde{O}=O-f$. Since $\langle O-f\rangle = \langle O \rangle $, which is independent of the neural network parameters, this constant term can be omitted. Consequently, the loss function simplifies to:
\begin{equation}
    L(w) = \frac{1}{N} \sum_{i=1}^N \left(O(\phi_i)-f(\phi_i) \right)^2 , 
    \label{Eq:Loss}
\end{equation}
where $w$ represents all the network parameters, such as $W_i$ and $b_i$, and $N$ denotes the number of training configurations.

Since the number of training configurations is finite (typically on the order of thousands in lattice calculations) and neural networks usually have a large number of parameters, overfitting is a common issue. In our case, overfitting imposes $O(\phi_i) = f(\phi_i)$ for every training configuration, which leads the sample average to be $\bar{f} \approx \bar{O}$, which is generally not zero. This issue was recognized in~\cite{mu}, where the authors suggested introducing an auxiliary parameter $\mu$:
\begin{equation}
    L(w, \mu) = \frac{1}{N} \sum_{i=1}^N \left(O(\phi_i)-f(\phi_i) - \mu \right)^2. \label{Eq:Loss_mu}
\end{equation}
In this way, when overfitting occurs, the auxiliary parameter $\mu$ may absorb the sample average of $O$, $\bar{O}$, ensuring that $\bar{f}$ remains close to zero. Note that during training using gradient descent, the auxiliary parameter $\mu$ is updated along the network parameters $w$. However, since $\mu$ is not used for evaluating observables, the new observable $\tilde{O}=O-f$ remains unbiased.

Therefore, in this proceeding, neural networks are optimized by minimizing the loss function in Eq.~(\ref{Eq:Loss_mu}), with the
\textit{ADAM} optimizer~\cite{ADAM} with exponentially decaying learning rate from $10^{-3}$ to $10^{-6}$.

\section{Scalar field theory}

\begin{figure*}[t!]
\centering
\includegraphics[width=0.60\textwidth]{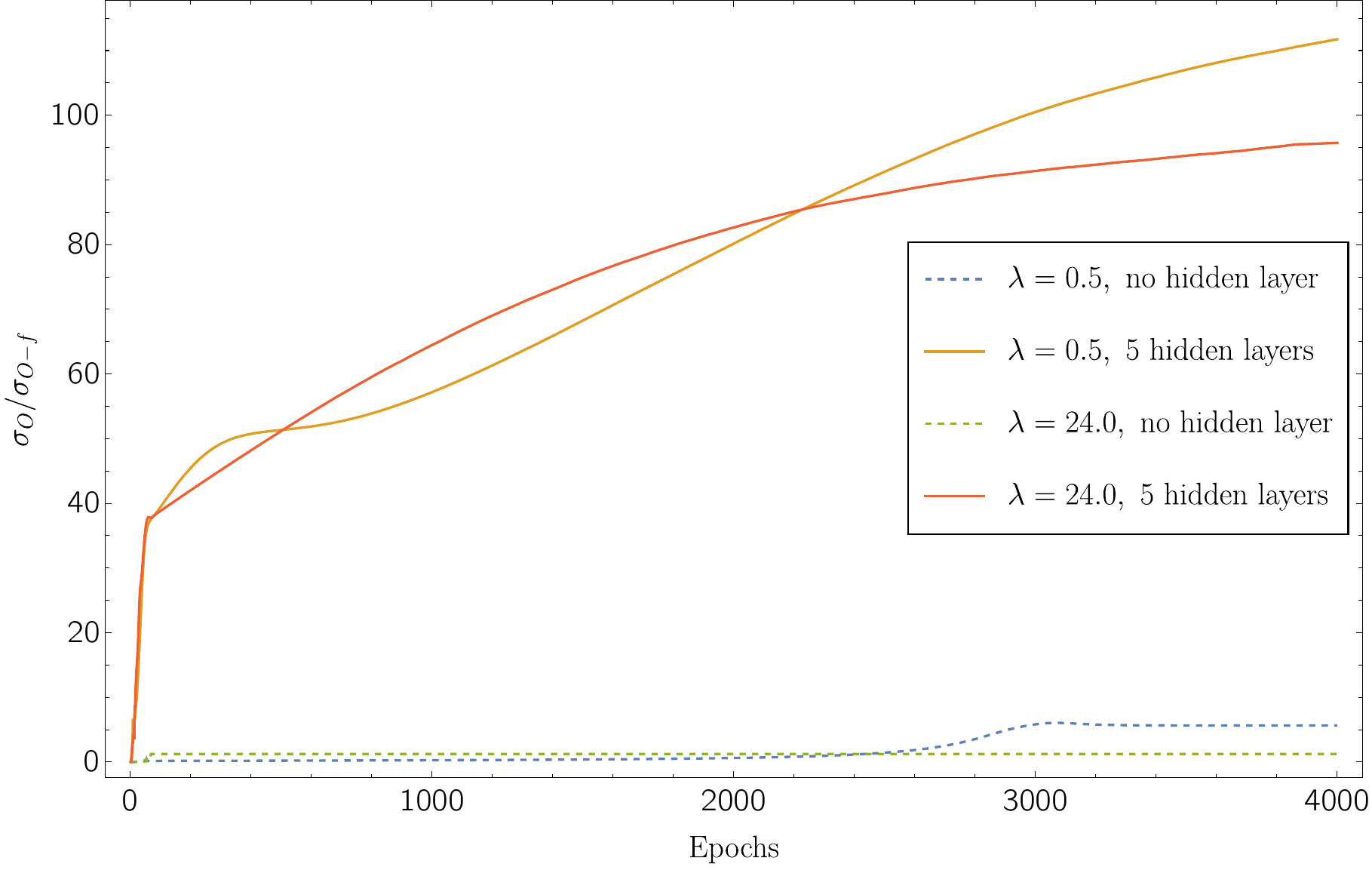}
\caption{Training histories of the small ($\lambda=0.5$) and large ($\lambda=24.0$) couplings on $20 \times 20$ lattice with $m^2=0.1$ are shown. The figure displays the improvement of standard deviation using control variates relative to the raw observabale $(\sigma_{\rm Raw}/\sigma_{\rm CV})$. The dashed line corresponds to the zero hidden layers (linear transformation), while the solid line represents the result with 5 hidden layers, each containing 4 neurons. $10^4$ configurations are used for training the network, and $10^3$ configurations are employed to estimate the standard deivation.}
\label{fig1}
\end{figure*}

In this section, we will utilize $1+1$ dimensional scalar field theory, which has served as a testbed for addressing signal-to-noise problems~\cite{CV, Scalar, Contour}.

On $L_0\times L_1$ lattice, the lattice action is written as
\begin{equation}
    S= \sum_{x, \mu} \left( \frac{(\phi_{x+\mu} - \phi_x)^2}{2} + \frac{m^2}{2}\phi_x^2 + \frac{\lambda}{4!}\phi_x^4  \right),
\end{equation}
where $\mu=0,\,1$ denotes the direction and and $x=(x_0,x_1)$. In this work, both small and large coupling cases are considered, with $\lambda=0.5,\,24.0$, respectively. The observable considered here is the two-point correlator at momentum $p=0$:
\begin{equation}
    O(t) = \frac{1}{L_0} \sum_{y_0} \left[ \left(  \frac{1}{L_1} \sum_{x_1} \phi_{y_0+t, x_1} \right) \left(  \frac{1}{L_1} \sum_{y_1} \phi_{y_0, y_1} \right) \right].
\end{equation}

Fig.~\ref{fig1} illustrates the training history of control variates on a $20\times20$ lattice for the correlator at the midpoint, $t=L_0/2$, where the signal-to-noise ratio is the worst. The results show that with zero hidden layer, corresponding to linear transformations, the control variates method still provides an improvement for the small coupling case, as discussed in~\cite{CV}, while it does not perform well for the large coupling case. However, by adding hidden layers, the improvement is in the order of hundreds.

\subsection{Transfer learning}

In this subsection, we discuss the improvement in the estimation of mass using a $40\times10$ lattice with $m^2=0.01$ and $\lambda=0.1$. Note that $L_2$ regularization is implemented and the loss function is given by:
\begin{equation}
    L(w, \mu) = \frac{1}{N} \sum_{i=1}^N \left(O(\phi_i)-f(\phi_i) - \mu \right)^2 + \delta \sum_w w^2, \label{Eq:Loss_L2}
\end{equation}
where $\delta$ denotes the regularization strength.

In the large $t$ limit, the correlator should follow the exponential fit:
\begin{equation}
    C(t) = B\left( {\rm e}^{-m t} + {\rm e}^{m (L_0 - t)} \right).
    \label{Eq:Fit}
\end{equation}

\begin{table}[b]
\centering
\begin{tabular}{|| c | c | c | c ||}
    \hline
     & $B$  & $m$   &  $\chi^2/{\rm dof}$ \\
    \hline
    Raw  & 0.000112(18) & 0.1935(40) & 0.55 \\
    \hline 
    CV at $t=20$  &  0.0001191(51) & 0.1920(13) & 0.29 \\
    \hline
    Transfer learning & 0.0001096(7) & 0.1938(2) & 0.91 \\
    \hline
\end{tabular}
    \caption{The fitted values of correlators in Fig.~\ref{fig2}.} \label{Table}
\end{table}

\begin{figure*}[t!]
\centering
\includegraphics[width=0.98\textwidth]{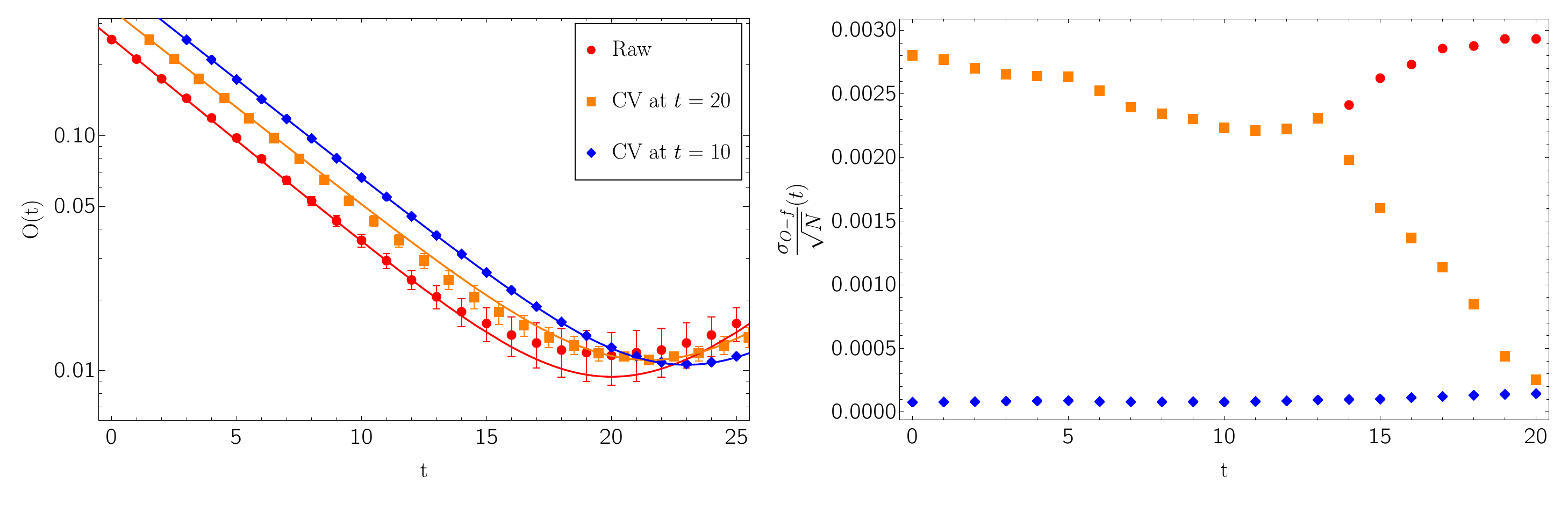}
\caption{Correlation functions with $m^2=0.01$ and $\lambda=0.1$ on $40 \times 10$ lattice are shown. The raw result and the result with control variates are shown. 
A total of $2\times10^3$ samples are used, with $10^3$ samples reserved for training the control variate, and the entire samples used for estimating the observables.
The left plot shows the correlation functions along with their fitting curves, as described in Eq.~(\ref{Eq:Fit}), with the results horizontally shifted for better visualization. The right plot displays the errors presented in the left plot.
}
\label{fig2}
\end{figure*}

Fig.~\ref{fig2} compares three correlator fits using different estimators: the naive observable, the improved observable with a control variate at $t=L_0/2=20$, and the improved estimator at all points using control variates. The fitting results are presented in Table~\ref{Table}. Note that a control variate at $t=20$  can be applied to other $t$-values if it helps reduce the variance. Therefore, it is utilized near $t=20$, as shown in the right panel of Fig.~\ref{fig2}.

Note that for the improved estimator at all points using control variates, the transfer learning technique is employed. In general, the transfer learning method~\cite{Transfer} leverages learned network parameters to optimize other networks. In our case, for example, if a network is found for the control variate at $t=10$, the network at 
$t=10$ can be used to initialize networks for control variates at 
$t=9$ and $t=11$. 
This approach significantly reduces the training time. Moreover, the right panel of Fig.~\ref{fig2} shows that the control variate trained with the transfer learning technique at $t=20$ outperforms the control variates trained from random initialization.

\section{Abelian gauge theory} \label{Sec:Gauge}

In this section, we consider 2-dimensional $U(1)$ gauge theory. The action is written as
\begin{equation}
    S= \beta \sum_{x} \cos(\theta^P_x),
\end{equation}
where $\theta^P$ denotes the plaquette angle, $\theta^L_1+\theta^L_2-\theta^L_3-\theta^L_4$ in terms of link angles. The observable is the Wilson loop:
\begin{equation}
    O(A) = \prod_{x\in A} e^{i \theta_x}. \label{Eq:Wilsonloop}
\end{equation}
Note that the Wilson loop is not averaged over various shapes or locations with the same size. While this is one of the simplest methods to reduce errors, it is avoided here to preserve the factorization property of a single Wilson loop.

With the open boundary condition and maximal tree gauge fixing, one can rewrite the integration measure of the partition function in terms of plaquettes:
\begin{equation}
    Z=\int D\theta^p \, \rm{exp}\left( \beta \sum_x \cos(\theta^P_x) \right) ,
\end{equation}
which have been used to study signal-to-noise problems~\cite{Contour} and sign problems~\cite{Bound}.

By using the factorization property of the action and the observable, one can construct a control variate as following: 
\begin{equation}
    f = \prod_{x \in A}{\rm e}^{i \theta^P_x} -  \prod_{x \in A} \left({\rm e}^{i \theta^P_x} - f_1(\theta^P_x)\right) \text{\, where\, } f_1(\theta^p) = \frac{dg}{d\theta^P} - g \frac{dS_1}{d\theta^P} . \label{Eq:2D_ansatz}
\end{equation}
Here, $S_1(\theta^P) = \beta (1-\cos \theta^p)$ is the action for a single plaquette. Note that since the observable in Eq.~(\ref{Eq:Wilsonloop}) is complex while its average is real, $f_1$ must generally be complex. Consequently, $g$ is parametrized by two networks, $g_r$ and $g_i$, such that $g(\theta^P) = g_r(\theta^P) + ig_i(\theta^P)$.

\begin{figure*}[t!]
\centering
\includegraphics[width=0.98\textwidth]{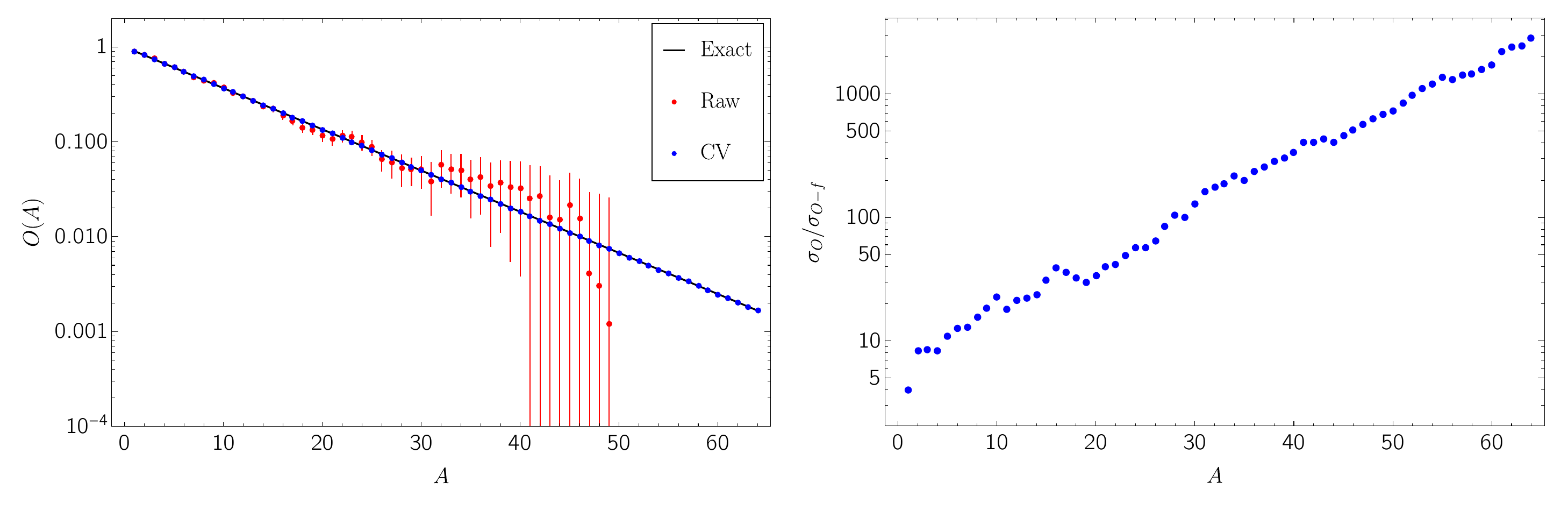}
\caption{Wilson loops on $8\times 8$ lattice, with the coupling $\beta=5.555$ are shown. The left panel represents the expectation values of Wilson loops with $1\sigma$ error bars, while the right plot illustrates the improvement in standard deviation  achieved using control variates, expressed as the ratio $(\sigma_{\rm Raw}/\sigma_{\rm CV})$.
}
\label{fig3}
\end{figure*}

Fig.~\ref{fig3} illustrates the error reduction achieved using neural control variates. Since two dimensional gauge theories are exactly solvable, the exact expectation values, $\left(I_1(\beta)/I_0(\beta)\right)^A$, are plotted for comparison. For each area, the two neural networks for $g$ have 4 hidden layers with 16 neurons each. To ensure that $f$ respects the periodic boundary condition to satisfy the Stein's identity, Eq.~(\ref{Eq:Stein}), the input of $g$ is is constructed from trigonometric functions of the plaquette angles $\theta_P$. Fig.~\ref{fig3} demonstrates that while the raw observable suffers from the signal-to-noise problem as the area increases, this issue is mitigated by the control variates, resulting in exponential error reduction, as shown in the right panel.

\section{Perfect control variates and constructions of control variates} \label{Sec:PCV}

One of the key properties of the control variates method is its potential to completely eliminate errors by identifying the perfect control variate $f_P$:
\begin{equation}
    f_P = O - \langle O \rangle. \label{Eq:PCV}
\end{equation}
Therefore, to evaluate the effectiveness of control variates in reducing variance, it is essential to determine whether the construction includes the perfect control variate or to assess how closely the constructed control variate approximates the ideal one.

There are multiple ways to construct control variates using the Stein's method. However, given that most theories respect translational symmetry, two general constructions can be considered. The first construction is 
\begin{equation}
    f(\phi) = \sum_i \left( \frac{\partial g}{\partial \phi_i} - g \frac{\partial S}{\partial \phi_i} \right) \; {\rm where} \; g:\mathbb R^V \rightarrow \mathbb R,  \label{Eq:CV_Gradient}
\end{equation}
Note that the scalar function $g$ must be translationally invariant to impose translational symmetry on the control variate $f$. Another construction is using a vector-valued function $g$, as shown in Eq.~(\ref{Eq:CV_Divergence}).
Note that $g$ must be translationally covariant for the control variate $f$ to be translationally invariant, as discussed in Section~\ref{Sec:Sym}.

Ref.~\cite{FCV} discussed that Eq.~(\ref{Eq:CV_Divergence}), the construction of control variates with a vector-valued function $g$, can generate any possible control variates through topological arguments. However, the topological argument does not address whether Eq.~(\ref{Eq:CV_Gradient}), the construction with a scalar function, can generate every possible control variate. If this is the case, it would be easier to find better control variates, as it would allow for a more restricted search space.

Unfortunately, the example in Section~(\ref{Sec:Gauge}) demonstrates that Eq.~(\ref{Eq:CV_Gradient}) might not generally include the perfect control variate. Consider the 2-area Wilson loop, where its plaquette angles are parametrized by $x$ and $y$. In this case, finding the perfect control variate corresponds to solving the following differential equation:
\begin{equation}
    \frac{\partial g}{\partial x} + \frac{\partial g}{\partial y} - \beta \sin(x) g - \beta \sin(y) g = \exp(i(x+y)) - \left(\frac{I_1(\beta)}{I_0(\beta)}\right)^2.
\end{equation}
One can easily find its general solution by changing variables to $x+y$ and $x-y$:
\begin{equation}
    g(x, y) = {\rm e}^{-2 \beta \cos\left(\frac{x+y}{2}\right) \cos\left(\frac{x-y}{2}\right)} \left( \int^{\frac{x+y}{2}} du \, {\rm e}^{2\beta \cos\left(u \right) \cos\left(\frac{x-y}{2}\right)} \left( {\rm e}^{2iu}-\left(\frac{I_1(\beta)}{I_0(\beta)}\right)^2 \right) + C\left(\frac{x-y}{2} \right) \right) ,
\end{equation}
where the function $C$ depends on the initial or boundary condition. It is easy to verify that this general solution does not have a periodic solution.

\section{Conclusion and outlook} \label{Sec:Conclusion}

In this work, we showed that the control variates method offers a promising way to mitigate signal-to-noise problems and it can be naturally combined with neural networks. We applied this method to $1+1$ dimensional scalar field theory and two dimensional $U(1)$ gauge theory to illustrate its potential for reducing signal-to-noise problems. Symmetries were imposed on the control variates to improve neural network performance.

In Section~\ref{Sec:PCV}, we discussed that the control variates derived from vector-valued functions are universal, whereas those from scalar-valued functions are not. 
However, finding effective control variates through training remains challenging, similar to the way solutions to differential equations are proven to exist, but their construction remains a separate challenge.
Using the Wilson loop example (Eq.~(\ref{Eq:2D_ansatz})), we showed that good control variates can be identified by leveraging the properties of the theory.
Future studies should explore other constructions, while considering specific theories and observables for better training.
Lastly, the application of the control variates method to more complex theories remains an important direction, and we are currently working on extending this approach to three-dimensional gauge theories.

\acknowledgments

This work was done in collaboration with Paulo Bedaque and Srijit Paul. The author expresses gratitude to Scott Lawrence for invaluable discussions. This work was supported in part by the U.S. Department of Energy, Office of Nuclear Physics under Award Number(s) DE-SC0021143, and DE-FG02-93ER40762.


\begin{thebibliography}{99}

\bibitem{CV}
T.~Bhattacharya, S.~Lawrence, and J.-S.~Yoo, 
\emph{Control variates for lattice field theory},
\href{https://journals.aps.org/prd/abstract/10.1103/PhysRevD.109.L031505}{Phys. Rev. D \textbf{109}, (2024) 3, L031505}
[\href{https://arxiv.org/abs/2307.14950}{2307.14950}].

\bibitem{Stein}
C.~Stein, 
\emph{A bound for the error in the normal approximation to the distribution of a sum of dependent random variables},
\href{https://projecteuclid.org/ebooks/berkeley-symposium-on-mathematical-statistics-and-probability/Proceedings-of-the-Sixth-Berkeley-Symposium-on-Mathematical-Statistics-and/chapter/A-bound-for-the-error-in-the-normal-approximation-to/bsmsp/1200514239}{Proceedings of the Sixth Berkeley Symposium on Mathematical Statistics and
Probability (Univ. California, Berkeley, Calif., 1970/1971), Vol. II: Probability theory, 583–602}.

\bibitem{FCV}
S.~Lawrence,
\emph{Schwinger-Dyson control variates for lattice fermions},
\href{https://arxiv.org/abs/2404.10707}{2404.10707}.


\bibitem{Oh}
P.~F.~Bedaque, and H.~Oh., 
\emph{Leveraging neural control variates for enhanced precision in lattice field theory},
\href{https://journals.aps.org/prd/abstract/10.1103/PhysRevD.109.094519}{Phys. Rev. D \textbf{109}, (2024) 094519}
[\href{https://arxiv.org/abs/2312.08228}{2312.08228}].

\bibitem{MLP}
K.~Hornik, M.~Stinchcombe, and H.~White,
\emph{Multilayer feedforward networks are universal approximators},
\href{https://www.sciencedirect.com/science/article/abs/pii/0893608089900208}{Neural Networks \textbf{2},
359 (1989)}.

\bibitem{MLP2}
G.~Cybenko, 
\emph{Approximation by superpositions of a sigmoidal function},
\href{https://doi.org/10.1007/BF02551274}{Math. Control Signal Systems \textbf{2}, 303–314 (1989)}.

\bibitem{mu}
R.~Wan, M.~Zhong, H.~Xiong, and Z.~Zhu,
\emph{Neural control variates for Monte Carlo variance reduction},
\href{https://doi.org/10.1007/978-3-030-46147-8_32}{In: Brefeld, U., Fromont, E., Hotho, A., Knobbe, A., Maathuis, M., Robardet, C. (eds) Machine Learning and Knowledge Discovery in Databases. ECML PKDD 2019. Lecture Notes in Computer Science(), vol 11907. Springer, Cham.}

\bibitem{ADAM}
D.~Kingma, and J.~Ba, 
\emph{Adam: A Method for Stochastic Optimization},
\href{https://arxiv.org/abs/1412.6980}{in International Conference on Learning Representations (ICLR) (San Diega, CA, USA, 2015)}.

\bibitem{Transfer}
S.~Bozinovski, and A.~Fulgosi,
\emph{The influence of pattern similarity and transfer learning on the base perceptron training (original in Croatian)}, 
Proceedings of Symposium Informatica 3-121-5, Bled (1976).

\bibitem{Scalar}
S.~Lawrence, H.~Oh., and Y.~Yamauchi,
\emph{Lattice scalar field theory at complex coupling},
\href{https://journals.aps.org/prd/abstract/10.1103/PhysRevD.106.114503}{Phys. Rev. D \textbf{106}, (2022) 114503}
[\href{https://arxiv.org/abs/2205.12303}{2205.12303}].

\bibitem{Contour}
W.~Detmold, G.~Kanwar, M.~L.~Wagman, and N.~C.~Warrington,
\emph{Path integral contour deformations for noisy observables},
\href{https://journals.aps.org/prd/abstract/10.1103/PhysRevD.102.014514}{Phys. Rev. D \textbf{102}, (2020) 014514}
[\href{https://arxiv.org/abs/2003.05914}{2003.05914}].

\bibitem{Bound}
S.~Lawrence, and Y.~Yamauchi,
\emph{Lattice scalar field theory at complex coupling},
\href{https://journals.aps.org/prd/abstract/10.1103/PhysRevD.110.014508}{Phys. Rev. D \textbf{110}, (2024) 014508}
[\href{https://arxiv.org/abs/2311.13002}{2311.13002}].



\end{thebibliography}
\end{document}